\documentclass[12pt]{article}
\usepackage{latexsym,amsmath,amssymb,amsbsy,graphicx,epsf}
\newcommand{\bs}{\boldsymbol}

\title{ The Effect of Slope in the Casimir Rack Gear }
\author{
Yu.S.Voronina,\footnote{E-mail: voronina@hep.phys.msu.ru} \ \ 
P.K.Silaev
\\ \it Moscow State University, Physics Dept.  
}

\begin{document}
\maketitle
\begin{abstract}
The effect of slope for the rack gear in the massless scalar field model is considered. 
It appears, that
the slope of profile surfaces can essentially change the value of normal Casimir force,
whereas average value of tangential force remains almost unchanged. At the same time
we observe essential shift 
of maximum and minimum tangential force positions.  
\end{abstract}

\section{Introduction}

In recent years  the Casimir effect \cite{casimir} has been a subject of extremely active consideration.  Experimental \cite{exp1,exp2,ref1,ref2}, and theoretical results \cite{ref1,ref2} 
deal with the attractive and repulsive Casimir force, with the Casimir force between
separated bodies (including anisotropic atoms) and the Casimir effect for one isolated body, Casimir effect for various topology, geometry, type of boundary condition and curvature of surfaces. 
For the purposes of  construction of nanomechanical devices there are two the most important aspects of the Casimir effect. The first aspect is the positive sign of energy (repulsion \cite{rep1}) without intermediate dielectric fluid, and the second one is the value of tangential force.  We will concentrate on the second task.

\begin{figure}
\includegraphics{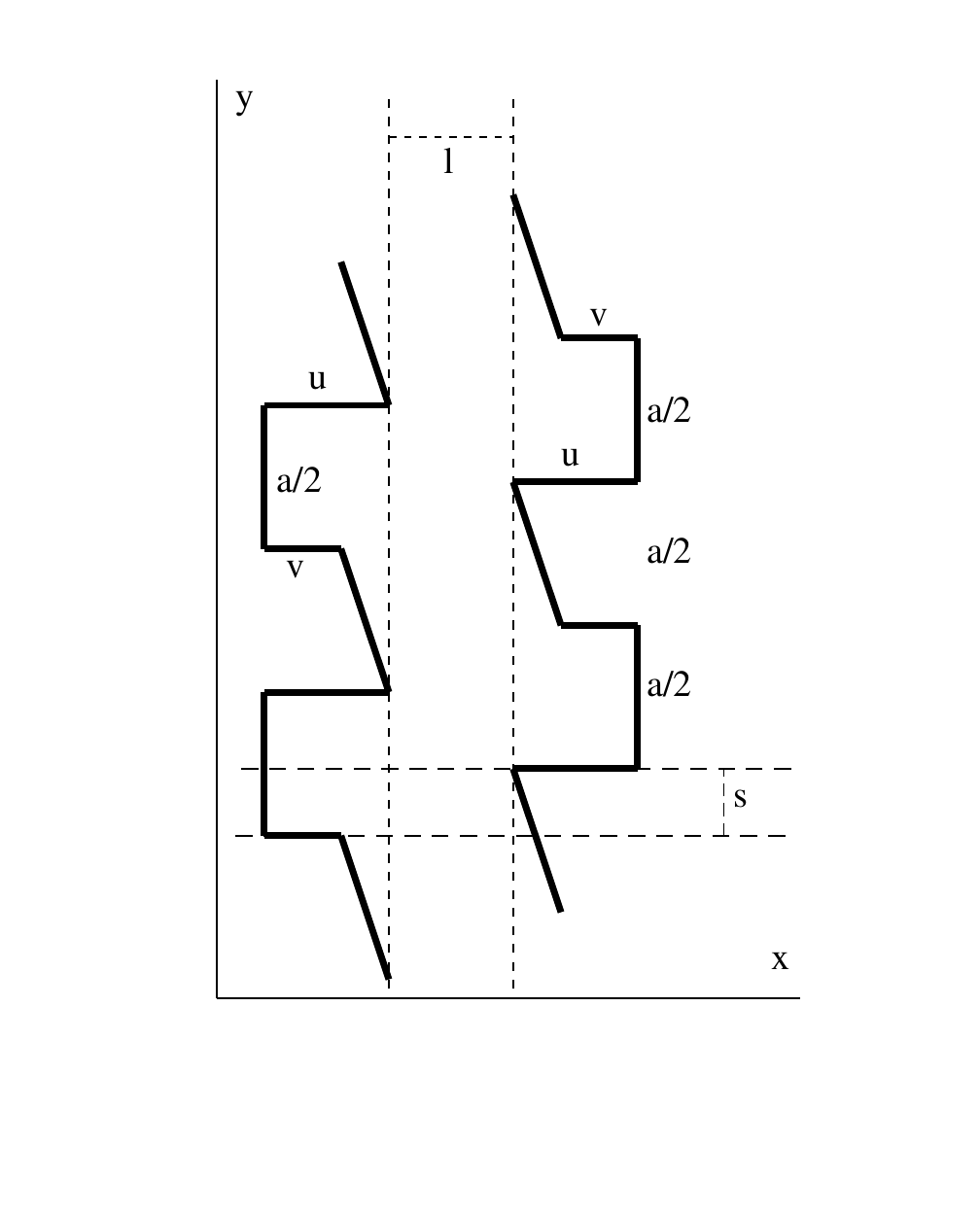}
\caption{System parameters: $a$ --- period, $u$ --- length of first profile edge, $v$ --- length of second profile edge, $s$ --- shift, $l$ --- width of the gap.
 }
\end{figure}

In this paper we consider the effect of slope for the rack gear. 
 For the sake of simplicity we study the case of scalar field, so one numerical calculation permits us to 
obtain  results   for both  two-dimensional 
and three dimensional geometry. 
The standard rack gear consists of  two profiled plates with identical period $a$. 
We assume, that the plates are parallel, whereas the surface of profiles has a slope 
 (see Fig 1. for geometry details and parameters). Three-dimensional model  implies additional dimension $z$, 
orthogonal to $x$ and $y$, with the translational invariance along $z$ axis. We assume the simplest 
zero boundary condition on the plate surface.  
  We start with the zero slope ($u$=$v$), and study the dependence of
both normal and tangential Casimir forces on the parameter $v$ for fixed $l$ and $u$. 
It is obvious, that  both forces will decrease for  decreasing $v$ (increasing slope),
 due to the  enlargement of  average distance between plates. 
But the  	type of the dependence  for normal and tangential forces can be different. Indeed,   
the density of normal force (normal force per unit length along the $y$ axis for two-dimensional case and
normal force corresponding to unit area at $y-z$ plane for three-dimensional case) depends essentially on
the distance between plates for all $y$ positions. At the same time the tangential force isn't equal to zero 
 only when this distance is not
constant, so for rack gear it is mostly side effect. For the geometry assumed the distance between 
profile edges with the length $u$  remains unchanged (it is equal to $l$), so the effect of slope for tangential force should be
less significant, then for normal force. 
We perform the direct numerical calculations to verify all these speculations.

\section{Calculation method}
As is well known, 
the simplest method to estimate  the Casimir force between two isolated bodies for the scalar field is to calculate numerically the Euclidian Green function for the surface conditions imposed \cite{fgg1,fgg2,fgg3}. 
So we'll find the solution to the equation 
\begin{equation}
\triangle G(\mu,\bs x,\bs y) - \mu^2 G(\mu,\bs x,\bs y) = \delta(\bs x-\bs y)
\end {equation}
with the boundary condition 
\begin{equation}
 G(\mu,\bs \xi,\bs x)=0, 
\end {equation}
where $\bs \xi$ lies on the plate surface. 

For any point $\bs x$ between  plates the vacuum expectation value of energy-momentum tensor density can be calculated as 
 $$\langle 0|T_{11}(\bs x)|0\rangle={1 \over \pi}\int\limits_0^\infty dq  \left( {1\over 2}\partial_{x_1}\partial_{y_1}-{1\over 2}\partial_{x_2}\partial_{y_2} -
{1\over 2} (q^2+m^2)  \right)\times$$
\begin{equation}
\times\left. G(\sqrt{q^2+m^2},\bs x,\bs y) \right|_{\bs y=\bs x} \label{n2d}
\end {equation}

\begin{equation}
 \langle 0|T_{12}(\bs x)|0\rangle= {1 \over \pi}\int\limits_0^\infty dq  \left( {1\over 2}
\partial_{x_1}\partial_{y_2} \right) 
\left. G(\sqrt{q^2+m^2},\bs x,\bs y) \right|_{\bs y=\bs x}
\end {equation}
for two-dimensional case. 
Here $m$ is the mass of the scalar field. 

Both expressions appears to be singular at  $\bs y=\bs x$ and should be renormalized. 
For two distant bodies the renormalization procedure is quite trivial --- it is   	sufficient to
subtract the same expression for Minkowski space.

The expression for three-dimensional case is almost identical due to translational invariance along
$z$ axis. One can perform Fourier transformation along $z$ axis and obtain two-dimensional task for each Fourier component. 
So the three-dimensional expressions take the form:
 $$\langle 0|T_{11}(\bs x)|0\rangle={1 \over 2\pi}\int\limits_0^\infty dq \;\sqrt{q^2+m^2} \left( {1\over 2}\partial_{x_1}\partial_{y_1}-{1\over 2}\partial_{x_2}\partial_{y_2} -
{1\over 2} (q^2+m^2)  \right)\times$$
\begin{equation}
\times\left. G(\sqrt{q^2+m^2},\bs x,\bs y) \right|_{\bs y=\bs x}
\end {equation}

\begin{equation}
 \langle 0|T_{12}(\bs x)|0\rangle= {1 \over 2\pi}\int\limits_0^\infty dq \;\sqrt{q^2+m^2} \left( {1\over 2}
\partial_{x_1}\partial_{y_2} \right) 
\left. G(\sqrt{q^2+m^2},\bs x,\bs y) \right|_{\bs y=\bs x} \label{t3d}
\end {equation}

Finally we should integrate renormalized expressions  (\ref{n2d})-(\ref{t3d}) along the $y$ axis from point
$\bs x_1=(x_0,y_0)$ to point  $\bs x_2=(x_0,y_0+a)$,
where                    $\bs x_1=(x_0,y_0)$ --- arbitrary point, that is placed  in the gap between plates. For instance, 
for (\ref{n2d}) (two-dimensional case, normal force):
\begin{equation}
F_{n}=\int\limits_0^a \langle 0|T_{11}^{ren}(x_0,y_0+w)|0\rangle \; dw.
\end {equation}
It is not the force density, but the force for one period;
 to obtain the force density  one should divide this result  by 
the period length $a$.

To calculate the difference between Green function and Green function for Minkowski space $G^{ren}(\mu,\bs x,\bs y)= G(\mu,\bs x,\bs y)-G^{(0)}(\mu,\bs x,\bs y)$ we use a slightly modified boundary-element method \cite{bem1,bem2,bem3}. 
This difference is equal to the solution of the homogeneous Helmholtz equation with the boundary condition $G^{ren}(\mu,\bs x,\bs \xi)  =- G^{0}(\mu,\bs x,\bs \xi) $, where $\bs \xi$ is positioned on the plate surface.
Instead of the standard spline approach for the given boundary element we use polynomial approximation, based on surrounding elements. 
It is similar, but not precisely equal to spline. 

We estimate calculation errors by increasing points density and subsequent comparison of results. 
The relative error appears to be about $10^{-2}$.

\section{Results and discussion}

We use the natural system of units $\hbar=c=1$ and choose the arbitrary geometry parameters $a=2$, $u=0.5$, $l=1$ and three values for
$v=0.5$ (zero slope),   $v=0.4$  (medium slope),   $v=0.3$  (large slope).  We should 
calculate
Green functions for all $\mu$ values, so we can obtain results for arbitrary mass of the
scalar field. But we restrict ourselves to the massless case $m=0$.  From equations (\ref{n2d})-(\ref{t3d}) one can easily conclude, that the mass
dependence can't drastically modify results,  so the type of dependence should remain unchanged.  
Direct calculations   	confirm this assumption.

\begin{figure}
\includegraphics{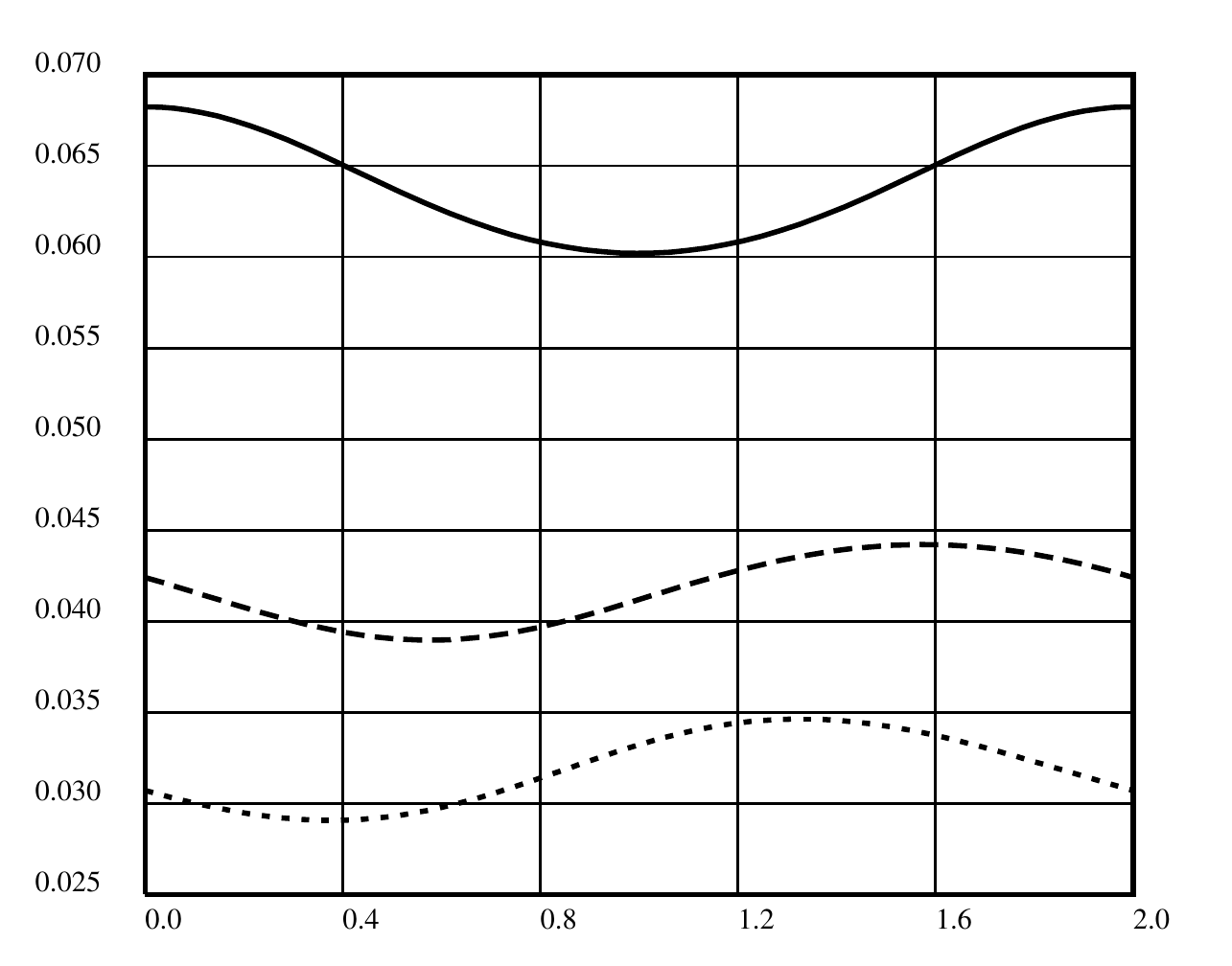}
\caption{Normal force for one period as the function of shift s for two-dimensional case.  
Solid line --- $v=0.5$, dashed line --- $v=0.4$, dotted line --- $v=0.3$. }
\end{figure}

\begin{figure}
\includegraphics{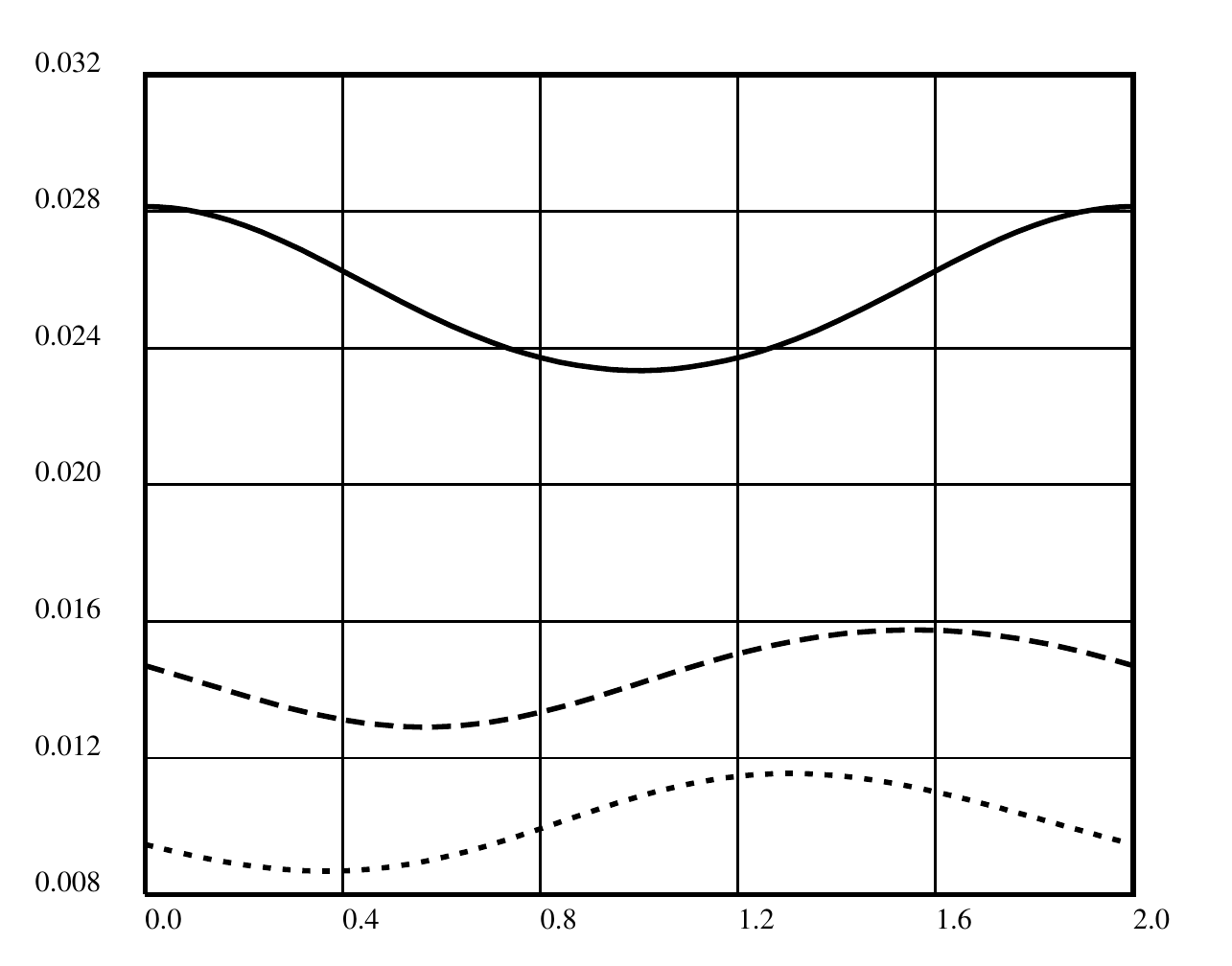}
\caption{Normal force for one period as the function of shift $s$ for three-dimensional case.   
Solid line --- $v=0.5$, dashed line --- $v=0.4$, dotted line --- $v=0.3$. }
\end{figure}

From Figs. 2 and 3 one can easily find, that the value of normal force essentially decreases for increasing slope 
for both  two- and three-dimensional cases. 
This result seems quite reasonable, if we take into account, that we change slope for fixed absolute gap width, 
whereas average gap width increases.  Moreover, this effect appears to be almost homogeneous: 
for different values of shift $s$ we observe almost identical slope dependence.   
It also should be noted, that there is no "phase shift" in the $s$-dependence, 
i.e. the position of maximum and minimum force values remains almost unchanged, at $s=0$ and $s=a/2$  correspondingly. 


\begin{figure}
\includegraphics{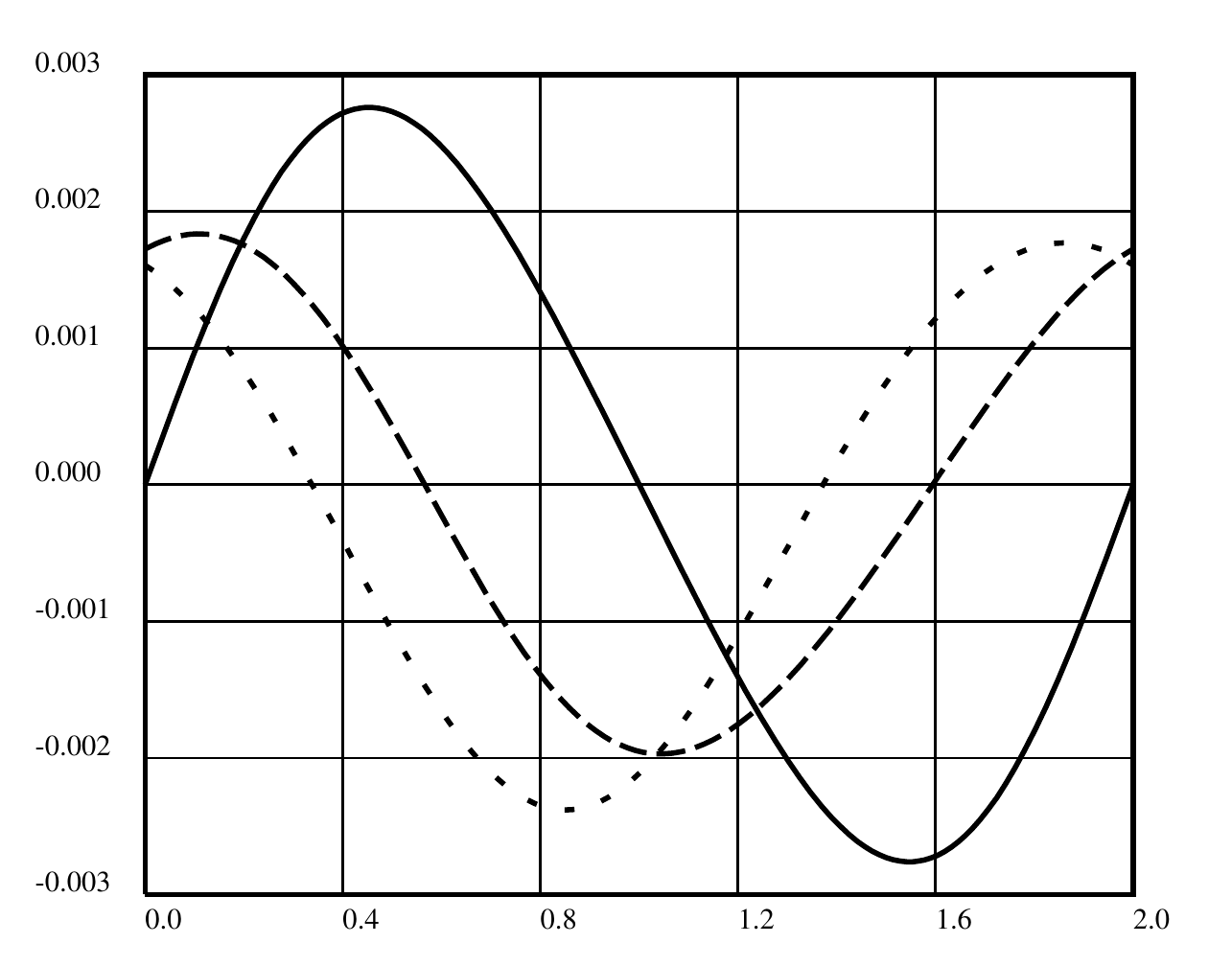}
\caption{Tangential force per one period as the function of shift s for two-dimensional case.  
Solid line --- $v=0.5$, dashed line --- $v=0.4$, dotted line --- $v=0.3$. }
\end{figure}

\begin{figure}
\includegraphics{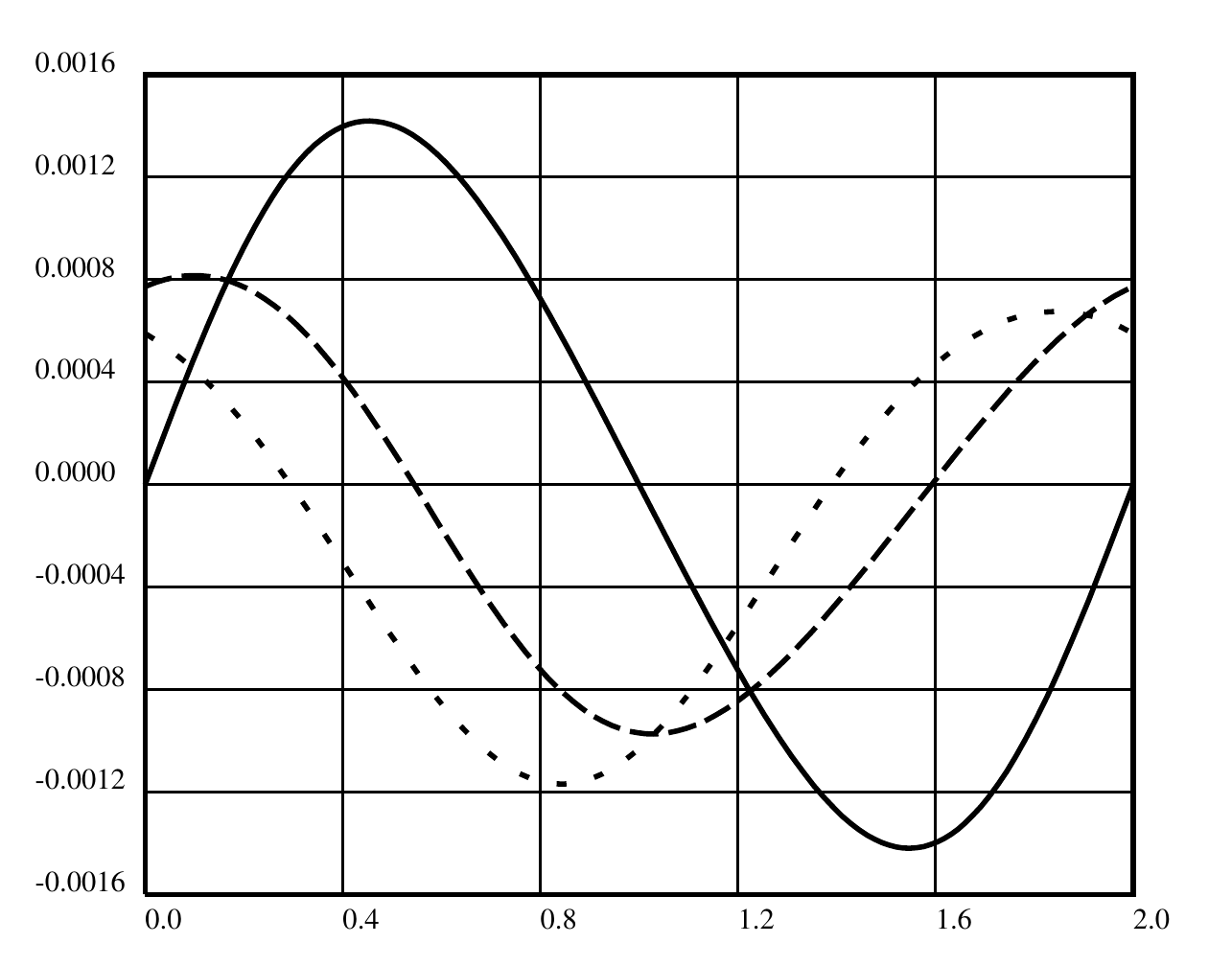}
\caption{Tangential force per one period as the function of shift s for three-dimensional case.  
Solid line --- $v=0.5$, dashed line --- $v=0.4$, dotted line --- $v=0.3$. }
\end{figure}

On the contrary, for the  tangential force (Figs. 4 and 5) we observe moderate decrease for increasing slope. 
Indeed, in two-dimensional case the normal force maximum for large slope ($v=0.3$) is about 45\%  of the maximum for zero slope ($v=0.5$), whereas the tangential force maximum  for large slope is about 85\%  of the maximum   for zero slope.
For three-dimensional case we obtain almost the same behavior, because the expressions  for three-dimensional case
differ from two-dimensional case only by one additional multiplier in the integral.

It also should be noted, that there is an essential "phase shift" in the $s$-dependence, i.e. the shift of the position for maximum and minimum force values  at various slopes. 
For instance, in the case of zero shift ($s=0$) we get nonzero tangential force for nonzero slopes, whereas  for
the zero slope  this force is absent due to obvious symmetry reasons. 
For the large slope this "phase shift" is almost equal to $a/2$ (about 40\% of the period $a$).    
In some sense the tangential force for the zero shift and nonzero slope is partly ``normal'' --- it can be interpreted  as the projection of attracting force between parallel sloped surfaces 
(this force is orthogonal to the surfaces)
on the  $y$ axis.

\section{Conclusions}

Direct numerical computations lead us to the conclusion, that (at least for the geometry considered)
the slope of profile surfaces can essentially change the value of normal Casimir force for the rack gear,
whereas tangential force remains almost unchanged, but exhibits essential "phase shift" --- shift of
maximum and minimum positions.   Both results are quite reasonable, because normal force is defined mostly by 
the width of the gap, whereas tangential force depends essentially on the change  of this width. 
For the geometry considered the derivative of the gap width  along $y$ axis depends on the slope, but average value of this derivative
remains almost unchanged.

It should be mentioned, that we consider the trivial case of zero boundary condition. 
For Neumann  or Robin  boundary conditions results may change drastically, 
because Green functions  at the presence of sharp edges (for instance, $\pi/2$ angles between flat surfaces)
essentially depend on the type of boundary condition considered.

\end{document}